\documentclass[10pt,a4paper,onecolumn]{article}
\usepackage{marginnote}
\usepackage{graphicx}
\usepackage{xcolor}
\usepackage{authblk,etoolbox}
\usepackage{titlesec}
\usepackage{calc}
\usepackage{tikz}
\usepackage{hyperref}
\hypersetup{colorlinks,breaklinks=true,
            urlcolor=[rgb]{0.0, 0.5, 1.0},
            linkcolor=[rgb]{0.0, 0.5, 1.0}}
\usepackage{caption}
\usepackage{tcolorbox}
\usepackage{amssymb,amsmath}
\usepackage{ifxetex,ifluatex}
\usepackage{seqsplit}
\usepackage{xstring}

\usepackage{float}
\let\origfigure\figure
\let\endorigfigure\endfigure
\renewenvironment{figure}[1][2] {
    \expandafter\origfigure\expandafter[H]
} {
    \endorigfigure
}

\usepackage{fixltx2e} 
\usepackage[
  backend=biber,
]{biblatex}
\bibliography{paper.bib}

\let\textttOrig=\texttt
\def\texttt#1{\expandafter\textttOrig{\seqsplit{#1}}}
\renewcommand{\seqinsert}{\ifmmode
  \allowbreak
  \else\penalty6000\hspace{0pt plus 0.02em}\fi}

\makeatletter
\let\href@Orig=\href
\def\href@Urllike#1#2{\href@Orig{#1}{\begingroup
    \def\Url@String{#2}\Url@FormatString
    \endgroup}}
\def\href@Notdoi#1#2{\def\tempa{#1}\def\tempb{#2}%
  \ifx\tempa\tempb\relax\href@Urllike{#1}{#2}\else
  \href@Orig{#1}{#2}\fi}
\def\href#1#2{%
  \IfBeginWith{#1}{https://doi.org}%
  {\href@Urllike{#1}{#2}}{\href@Notdoi{#1}{#2}}}
\makeatother

\newlength{\cslhangindent}
\newlength{\csllabelwidth}
\newenvironment{CSLReferences}[3] 
 {
  \setlength{\parindent}{0pt}
  \ifodd #1 \everypar{\setlength{\hangindent}{\cslhangindent}}\ignorespaces\fi
  \ifnum #2 > 0
  \setlength{\parskip}{#2\baselineskip}
  \fi
 }%
 {}
\usepackage{calc}

\usepackage[top=3.5cm, bottom=3cm, right=1.5cm, left=1.0cm,
            headheight=2.2cm, reversemp, includemp, marginparwidth=4.5cm]{geometry}

\titleformat{\section}
  {\normalfont\sffamily\Large\bfseries}
  {}{0pt}{}
\titleformat{\subsection}
  {\normalfont\sffamily\large\bfseries}
  {}{0pt}{}
\titleformat{\subsubsection}
  {\normalfont\sffamily\bfseries}
  {}{0pt}{}
\titleformat*{\paragraph}
  {\sffamily\normalsize}

\usepackage{fancyhdr}
\makeatletter
\let\ps@plain\ps@fancy
\fancyheadoffset[L]{4.5cm}
\fancyfootoffset[L]{4.5cm}

\definecolor{linky}{rgb}{0.0, 0.5, 1.0}

\newtcolorbox{repobox}
   {colback=red, colframe=red!75!black,
     boxrule=0.5pt, arc=2pt, left=6pt, right=6pt, top=3pt, bottom=3pt}

\newcommand{\ExternalLink}{%
   \tikz[x=1.2ex, y=1.2ex, baseline=-0.05ex]{%
       \begin{scope}[x=1ex, y=1ex]
           \clip (-0.1,-0.1)
               --++ (-0, 1.2)
               --++ (0.6, 0)
               --++ (0, -0.6)
               --++ (0.6, 0)
               --++ (0, -1);
           \path[draw,
               line width = 0.5,
               rounded corners=0.5]
               (0,0) rectangle (1,1);
       \end{scope}
       \path[draw, line width = 0.5] (0.5, 0.5)
           -- (1, 1);
       \path[draw, line width = 0.5] (0.6, 1)
           -- (1, 1) -- (1, 0.6);
       }
   }

\patchcmd{\@maketitle}{center}{flushleft}{}{}
\patchcmd{\@maketitle}{center}{flushleft}{}{}
\patchcmd{\@maketitle}{\LARGE}{\LARGE\sffamily}{}{}
\def\maketitle{{%
  
  \AB@maketitle}}
\makeatletter
\renewcommand\AB@affilsepx{ \protect\Affilfont}
\renewcommand\AB@affilnote[1]{{\bfseries #1}\hspace{3pt}}
\renewcommand{\affil}[2][]%
   {\newaffiltrue\let\AB@blk@and\AB@pand
      \if\relax#1\relax\def\AB@note{\AB@thenote}\else\def\AB@note{#1}%
        \setcounter{Maxaffil}{0}\fi
        \begingroup
        \let\href=\href@Orig
        \let\texttt=\textttOrig
        \let\protect\@unexpandable@protect
        \def\thanks{\protect\thanks}\def\footnote{\protect\footnote}%
        \@temptokena=\expandafter{\AB@authors}%
        {\def\\{\protect\\\protect\Affilfont}\xdef\AB@temp{#2}}%
         \xdef\AB@authors{\the\@temptokena\AB@las\AB@au@str
         \protect\\[\affilsep]\protect\Affilfont\AB@temp}%
         \gdef\AB@las{}\gdef\AB@au@str{}%
        {\def\\{, \ignorespaces}\xdef\AB@temp{#2}}%
        \@temptokena=\expandafter{\AB@affillist}%
        \xdef\AB@affillist{\the\@temptokena \AB@affilsep
          \AB@affilnote{\AB@note}\protect\Affilfont\AB@temp}%
      \endgroup
       \let\AB@affilsep\AB@affilsepx
}
\makeatother

\renewcommand\Affilfont{\sffamily\small\mdseries}
\ifnum 0\ifxetex 1\fi\ifluatex 1\fi=0 
  \usepackage[T1]{fontenc}
  \usepackage[utf8]{inputenc}

\else 
  \ifxetex
    \usepackage{mathspec}
    \usepackage{fontspec}

  \else
    \usepackage{fontspec}
  \fi
  \defaultfontfeatures{Ligatures=TeX,Scale=MatchLowercase}

\fi
\IfFileExists{upquote.sty}{\usepackage{upquote}}{}
\IfFileExists{microtype.sty}{%
\usepackage{microtype}
\UseMicrotypeSet[protrusion]{basicmath} 
}{}

\usepackage{hyperref}
\hypersetup{unicode=true,
            pdftitle={exoTEDRF: An EXOplanet Transit and Eclipse Data Reduction Framework},
            pdfborder={0 0 0},
            breaklinks=true}
\let\addcontentslineOrig=\addcontentsline
\def\addcontentsline#1#2#3{\bgroup
  \let\texttt=\textttOrig\addcontentslineOrig{#1}{#2}{#3}\egroup}
\let\markbothOrig\markboth
\def\markboth#1#2{\bgroup
  \let\texttt=\textttOrig\markbothOrig{#1}{#2}\egroup}
\let\markrightOrig\markright
\def\markright#1{\bgroup
  \let\texttt=\textttOrig\markrightOrig{#1}\egroup}

\IfFileExists{parskip.sty}{%
\usepackage{parskip}
}{
\setlength{\parindent}{0pt}
\setlength{\parskip}{6pt plus 2pt minus 1pt}
}
\providecommand{\tightlist}{%
  \setlength{\itemsep}{0pt}\setlength{\parskip}{0pt}}
\ifx\paragraph\undefined\else
\let\oldparagraph\paragraph
\renewcommand{\paragraph}[1]{\oldparagraph{#1}\mbox{}}
\fi
\ifx\subparagraph\undefined\else
\let\oldsubparagraph\subparagraph
\renewcommand{\subparagraph}[1]{\oldsubparagraph{#1}\mbox{}}
\fi

\title{exoTEDRF: An EXOplanet Transit and Eclipse Data Reduction
Framework}

        \author[1]{Michael Radica}
    
      \affil[1]{Trottier Institute for Research on Exoplanets (iREx),
Université de Montréal, Montréal, Canada}
  \date{\vspace{-7ex}}

\begin{document}
\maketitle

\marginpar{

  \begin{flushleft}
  \sffamily\small

  {\bfseries DOI:} \href{https://doi.org/DOI unavailable}{\color{linky}{DOI unavailable}}

  \vspace{2mm}

  {\bfseries Software}
  \begin{itemize}
    \setlength\itemsep{0em}
    \item \href{N/A}{\color{linky}{Review}} \ExternalLink
    \item \href{NO_REPOSITORY}{\color{linky}{Repository}} \ExternalLink
    \item \href{DOI unavailable}{\color{linky}{Archive}} \ExternalLink
  \end{itemize}

  \vspace{2mm}

  \par\noindent\hrulefill\par

  \vspace{2mm}

  {\bfseries Editor:} \href{https://example.com}{Warrick Ball} \ExternalLink \\
  \vspace{1mm}
    {\bfseries Reviewers:}
  \begin{itemize}
  \setlength\itemsep{0em}
    \item \href{https://github.com/Pending Reviewers}{@karllark, @mm-murphy}
    \end{itemize}
    \vspace{2mm}

  {\bfseries Submitted:} 30 April 2024\\
  {\bfseries Published:} N/A

  \vspace{2mm}
  {\bfseries License}\\
  Authors of papers retain copyright and release the work under a Creative Commons Attribution 4.0 International License (\href{http://creativecommons.org/licenses/by/4.0/}{\color{linky}{CC BY 4.0}}).

  \end{flushleft}
}

\hypertarget{summary}{%
\section{Summary}\label{summary}}

Since the start of science operations in July 2022, JWST has delivered
groundbreaking results on a regular basis. For spectroscopic studies of
exoplanet atmospheres the process of extracting robust and reliable
atmosphere spectra from the raw JWST observations is critical.
Especially as the field pushes to detect the signatures of secondary
atmospheres on rocky Earth-like planets it is imperative to ensure that
the spectral features that drive atmosphere inferences are robust
against the particular choices made during the data reduction process.

Here, I present the community with \texttt{exoTEDRF} (EXOplanet Transit
and Eclipse Data Reduction Framework; formerly known as
\texttt{supreme-SPOON}), an end-to-end pipeline for data reduction and
light curve analysis of time series observations (TSOs) of transiting
exoplanets with JWST. The pipeline is highly modular and designed to
produce reliable spectra from raw JWST exposures. \texttt{exoTEDRF}
(pronounced exo-tedorf) consists of four stages, each of which are
further subdivided into a series of steps. These steps can either be run
individually, for example in a Jupyter notebook, or via the command line
using the provided configuration files. The steps are highly tunable,
allowing full control over every parameter in the reduction. Each step
also produces diagnostic plots to allow the user to verify their results
at each intermediate stage, and compare outputs with other pipelines if
so desired. Finally, \texttt{exoTEDRF} has also been designed to be run
in ``batch'' mode: simultaneously running multiple reductions, each
tweaking a subset of parameters, to understand any impacts on the
resulting atmosphere spectrum.

\hypertarget{overview-of-exotedrf-stages}{%
\section{Overview of exoTEDRF
Stages}\label{overview-of-exotedrf-stages}}

Like similar pipelines (\texttt{Eureka!} (Bell et al., 2022),
\texttt{jwst} (Bushouse et al., 2022), etc.) \texttt{exoTEDRF} is
divided up into four major stages which are summarized below:

\begin{itemize}
\tightlist
\item
  Stage 1, Detector-level processing: Converts raw, 4D (integrations,
  groups, \(x\)-pixel, \(y\)-pixel) data frames to 3D (integrations,
  \(x\)-pixel, \(y\)-pixel) slope images. Steps include superbias
  subtractions, correction of 1/\(f\) noise, ramp fitting, etc.
\item
  Stage 2, Spectroscopic processing: Performs additional calibrations to
  prepare slope images for spectral extraction. Steps include, flat
  field correction, background subtraction, etc.
\item
  Stage 3, Spectral extraction: Extract the 2D stellar spectra from the
  3D slope images.
\item
  Stage 4, Light curve fitting: An optional stage for the fitting of
  extracted light curves.
\end{itemize}

In \texttt{exoTEDRF}, Stage 4 is an optional installation which is
currently built around the \texttt{exoUPRF} library (Radica, 2024), and
incorporates tools such as \texttt{ExoTiC-LD} (Grant \& Wakeford, 2022)
for the estimation of stellar limb darkening parameters. In certain
places (e.g., superbias subtraction, flat field correction),
\texttt{exoTEDRF} simply provides a wrapper around the existing
functionalities of the \texttt{jwst} package maintained by the Space
Telescope Science Institute.

\hypertarget{statement-of-need}{%
\section{Statement of Need}\label{statement-of-need}}

Data analysis is a challenging process that is encountered by all
observational studies. Ensuring that the resulting atmosphere spectra
are robust against particular choices made in the reduction process is
critical, especially as we push to characterize the atmospheres of small
rocky planets. The modularity and tunability of \texttt{exoTEDRF} make
it easy to run multiple reductions of a given dataset, and therefore
robustly ascertain whether the spectral features driving atmosphere
inferences are robust, or sensitive to the peculiarities of a given
reduction. Moreover, \texttt{exoTEDRF} has full support for TSOs with
NIRISS/SOSS (Albert et al., 2023), an observing mode which is
underserved by the current ecosystem of JWST reduction tools, including
being the only pipeline with the ability to run the \texttt{ATOCA}
extraction algorithm (Darveau-Bernier et al., 2022; Radica et al., 2022)
to explicitly model the SOSS order overlap.

\hypertarget{documentation}{%
\section{Documentation}\label{documentation}}

Documentation for \texttt{exoTEDRF}, including example notebooks, is
available at \url{https://exotedrf.readthedocs.io/en/latest/}.

\hypertarget{uses-of-exotedrf-in-current-literature}{%
\section{Uses of exoTEDRF in Current
Literature}\label{uses-of-exotedrf-in-current-literature}}

\texttt{exoTEDRF} (particularly in its previous life as
\texttt{supreme-SPOON}) has been widely applied to exoplanet TSOs. A
list of current literature which has made use of \texttt{exoTEDRF}
includes: A. D. Feinstein et al. (2023), Coulombe et al. (2023), Radica
et al. (2023), Albert et al. (2023), Lim et al. (2023), Radica et al.
(2024), Fournier-Tondreau et al. (2024), Benneke et al. (2024), and
Cadieux et al. (2024).

\hypertarget{future-developments}{%
\section{Future Developments}\label{future-developments}}

The current release of \texttt{exoTEDRF} (v2.0.0) currently supports the
reduction of TSOs observed with JWST NIRISS/SOSS as well as
NIRSpec/BOTS. Support for observations MIRI/LRS is in development and
will be added in the coming months. \texttt{exoTEDRF} has also been
applied to exoplanet observations from the Hubble Space Telescope using
the UVIS mode (Radica et al., 2024b, in prep). This functionality will
also be made available to the public.

Suggestions for additional features are always welcome!

\hypertarget{similar-tools}{%
\section{Similar Tools}\label{similar-tools}}

The following is a list of other open source pipelines tailored to
exoplanet observations with JWST, some of which general purpose, and
others which are more tailored to specific instruments:

\begin{itemize}
\tightlist
\item
  General purpose: \texttt{Eureka!} (Bell et al., 2022), \texttt{jwst}
  (Bushouse et al., 2022), \texttt{transitspectroscopy} (Nestor
  Espinoza, 2022)
\item
  NIRISS specific: \texttt{nirHiss} (A. Feinstein, 2022)
\item
  NIRCam specific: \texttt{tshirt} (Schlawin \& Glidic, 2022)
\item
  MIRI specific: \texttt{PACMAN} (Zieba \& Kreidberg, 2022),
  \texttt{ExoTiC-MIRI} (Grant et al., 2023)
\item
  NIRSpec specific: \texttt{ExoTiC-JEDI} (Alderson et al., 2022)
\end{itemize}

Packages like \texttt{exoplanet} (Foreman-Mackey et al., 2021),
\texttt{Eureka!} (Bell et al., 2022), and \texttt{juliet} (Néstor
Espinoza et al., 2019) also enable similar light curve fitting.

\hypertarget{acknowledgements}{%
\section{Acknowledgements}\label{acknowledgements}}

The foundations of \texttt{exoTEDRF} are built upon many wonderful
Python libraries, including \texttt{numpy} (Harris et al., 2020),
\texttt{scipy} (Virtanen et al., 2020), \texttt{astropy} (Astropy
Collaboration et al., 2018, 2013), and \texttt{matplotlib} (Hunter,
2007).

MR acknowledges funding from the Natural Sciences and Engineering
Research Council of Canada, the Fonds de Recherche du Québec -- Nature
et Technologies, and the Trottier Institute for Research on Exoplanets.
He would also like to thank the JWST Transiting Exoplanet Community
Early Release Science program for providing the forum where much of the
development of this pipeline occured, and in particular, Adina
Feinstein, Louis-Philippe Coulombe, Néstor Espinoza, and Lili Alderson
for many helpful conversations.

\hypertarget{references}{%
\section*{References}\label{references}}
\addcontentsline{toc}{section}{References}

\hypertarget{refs}{}
\begin{CSLReferences}{1}{0}
\leavevmode\hypertarget{ref-Albert2023}{}%
Albert, L., Lafrenière, D., René, Doyon, Artigau, É., Volk, K.,
Goudfrooij, P., Martel, A. R., Radica, M., Rowe, J., Espinoza, N., Roy,
A., Filippazzo, J. C., Darveau-Bernier, A., Talens, G. J.,
Sivaramakrishnan, A., Willott, C. J., Fullerton, A. W., LaMassa, S.,
Hutchings, J. B., \ldots{} Kaltenegger, L. (2023). {The Near Infrared
Imager and Slitless Spectrograph for the James Webb Space Telescope.
III. Single Object Slitless Spectroscopy}. \emph{Publications of the
Astronomical Society of the Pacific}, \emph{135}(1049), 075001.
\url{https://doi.org/10.1088/1538-3873/acd7a3}

\leavevmode\hypertarget{ref-jedi2022}{}%
Alderson, L., Grant, D., \& Wakeford, H. (2022).
\emph{{Exo-TiC/ExoTiC-JEDI: v0.1-beta-release}} (Version v0.1)
{[}Computer software{]}. Zenodo.
\url{https://doi.org/10.5281/zenodo.7185855}

\leavevmode\hypertarget{ref-astropy:2018}{}%
Astropy Collaboration, Price-Whelan, A. M., Sipőcz, B. M., Günther, H.
M., Lim, P. L., Crawford, S. M., Conseil, S., Shupe, D. L., Craig, M.
W., Dencheva, N., Ginsburg, A., VanderPlas, J. T., Bradley, L. D.,
Pérez-Suárez, D., de Val-Borro, M., Aldcroft, T. L., Cruz, K. L.,
Robitaille, T. P., Tollerud, E. J., \ldots{} Astropy Contributors.
(2018). {The Astropy Project: Building an Open-science Project and
Status of the v2.0 Core Package}. \emph{The Astronomical Journal},
\emph{156}(3), 123. \url{https://doi.org/10.3847/1538-3881/aabc4f}

\leavevmode\hypertarget{ref-astropy:2013}{}%
Astropy Collaboration, Robitaille, T. P., Tollerud, E. J., Greenfield,
P., Droettboom, M., Bray, E., Aldcroft, T., Davis, M., Ginsburg, A.,
Price-Whelan, A. M., Kerzendorf, W. E., Conley, A., Crighton, N.,
Barbary, K., Muna, D., Ferguson, H., Grollier, F., Parikh, M. M., Nair,
P. H., \ldots{} Streicher, O. (2013). {Astropy: A community Python
package for astronomy}. \emph{Astronomy \& Astrophysics}, \emph{558},
A33. \url{https://doi.org/10.1051/0004-6361/201322068}

\leavevmode\hypertarget{ref-bell_eureka_2022}{}%
Bell, T., Ahrer, E.-M., Brande, J., Carter, A., Feinstein, A., Caloca,
G., Mansfield, M., Zieba, S., Piaulet, C., Benneke, B., Filippazzo, J.,
May, E., Roy, P.-A., Kreidberg, L., \& Stevenson, K. (2022). {Eureka!:
An End-to-End Pipeline for JWST Time-Series Observations}. \emph{Journal
of Open Source Software}, \emph{7}(79), 4503.
\url{https://doi.org/10.21105/joss.04503}

\leavevmode\hypertarget{ref-Benneke2024}{}%
Benneke, B., Roy, P.-A., Coulombe, L.-P., Radica, M., Piaulet, C.,
Ahrer, E.-M., Pierrehumbert, R., Krissansen-Totton, J., Schlichting, H.
E., Hu, R., Yang, J., Christie, D., Thorngren, D., Young, E. D.,
Pelletier, S., Knutson, H. A., Miguel, Y., Evans-Soma, T. M., Dorn, C.,
\ldots{} Allart, R. (2024). {JWST Reveals CH\(_4\), CO\(_2\), and
H\(_2\)O in a Metal-rich Miscible Atmosphere on a Two-Earth-Radius
Exoplanet}. \emph{arXiv e-Prints}, arXiv:2403.03325.
\url{https://doi.org/10.48550/arXiv.2403.03325}

\leavevmode\hypertarget{ref-bushouse_howard_2022_7038885}{}%
Bushouse, H., Eisenhamer, J., Dencheva, N., Davies, J., Greenfield, P.,
Morrison, J., Hodge, P., Simon, B., Grumm, D., Droettboom, M., Slavich,
E., Sosey, M., Pauly, T., Miller, T., Jedrzejewski, R., Hack, W., Davis,
D., Crawford, S., Law, D., \ldots{} Jamieson, W. (2022). \emph{JWST
calibration pipeline} (Version 1.7.0). Zenodo; Zenodo.
\url{https://doi.org/10.5281/zenodo.7038885}

\leavevmode\hypertarget{ref-Cadieux2024}{}%
Cadieux, C., Doyon, R., MacDonald, R. J., Turbet, M., Artigau, É., Lim,
O., Radica, M., Fauchez, T. J., Salhi, S., Dang, L., Albert, L.,
Coulombe, L.-P., Cowan, N. B., Lafrenière, D., L'Heureux, A., Piaulet,
C., Benneke, B., Cloutier, R., Charnay, B., \ldots{} Valencia, D.
(2024). {Transmission Spectroscopy of the Habitable Zone Exoplanet LHS
1140 b with JWST/NIRISS}. \emph{arXiv e-Prints}, arXiv:2406.15136.
\url{https://doi.org/10.48550/arXiv.2406.15136}

\leavevmode\hypertarget{ref-Coulombe2023}{}%
Coulombe, L.-P., Benneke, B., Challener, R., Piette, A. A. A., Wiser, L.
S., Mansfield, M., MacDonald, R. J., Beltz, H., Feinstein, A. D.,
Radica, M., Savel, A. B., Dos Santos, L. A., Bean, J. L., Parmentier,
V., Wong, I., Rauscher, E., Komacek, T. D., Kempton, E. M.-R., Tan, X.,
\ldots{} Wheatley, P. J. (2023). {A broadband thermal emission spectrum
of the ultra-hot Jupiter WASP-18b}. \emph{Nature}, \emph{620}(7973),
292--298. \url{https://doi.org/10.1038/s41586-023-06230-1}

\leavevmode\hypertarget{ref-Darveau-Bernier2022}{}%
Darveau-Bernier, A., Albert, L., Talens, G. J., Lafrenière, D., Radica,
M., Doyon, R., Cook, N. J., Rowe, J. F., Allart, R., Artigau, É.,
Benneke, B., Cowan, N., Dang, L., Espinoza, N., Johnstone, D.,
Kaltenegger, L., Lim, O., Pauly, T., Pelletier, S., \ldots{} Turner, J.
D. (2022). {ATOCA: an Algorithm to Treat Order Contamination.
Application to the NIRISS SOSS Mode}. \emph{Publications of the
Astronomical Society of the Pacific}, \emph{134}(1039), 094502.
\url{https://doi.org/10.1088/1538-3873/ac8a77}

\leavevmode\hypertarget{ref-espinoza_nestor_2022_6960924}{}%
Espinoza, Nestor. (2022). \emph{TransitSpectroscopy} (Version 0.3.11)
{[}Computer software{]}. Zenodo.
\url{https://doi.org/10.5281/zenodo.6960924}

\leavevmode\hypertarget{ref-espinoza_juliet_2019}{}%
Espinoza, Néstor, Kossakowski, D., \& Brahm, R. (2019). {juliet: a
versatile modelling tool for transiting and non-transiting exoplanetary
systems}. \emph{Monthly Notices of the Royal Astronomical Society},
\emph{490}(2), 2262--2283. \url{https://doi.org/10.1093/mnras/stz2688}

\leavevmode\hypertarget{ref-nirhiss2022}{}%
Feinstein, A. (2022). nirHiss. In \emph{GitHub repository}. GitHub.
\url{https://github.com/afeinstein20/nirhiss}

\leavevmode\hypertarget{ref-Feinstein2023}{}%
Feinstein, A. D., Radica, M., Welbanks, L., Murray, C. A., Ohno, K.,
Coulombe, L.-P., Espinoza, N., Bean, J. L., Teske, J. K., Benneke, B.,
Line, M. R., Rustamkulov, Z., Saba, A., Tsiaras, A., Barstow, J. K.,
Fortney, J. J., Gao, P., Knutson, H. A., MacDonald, R. J., \ldots{}
Zhang, X. (2023). {Early Release Science of the exoplanet WASP-39b with
JWST NIRISS}. \emph{Nature}, \emph{614}(7949), 670--675.
\url{https://doi.org/10.1038/s41586-022-05674-1}

\leavevmode\hypertarget{ref-exoplanet:joss}{}%
Foreman-Mackey, D., Luger, R., Agol, E., Barclay, T., Bouma, L. G.,
Brandt, T. D., Czekala, I., David, T. J., Dong, J., Gilbert, E. A.,
Gordon, T. A., Hedges, C., Hey, D. R., Morris, B. M., Price-Whelan, A.
M., \& Savel, A. B. (2021). {exoplanet: Gradient-based probabilistic
inference for exoplanet data \& other astronomical time series}.
\emph{arXiv e-Prints}, arXiv:2105.01994.
\url{http://arxiv.org/abs/2105.01994}

\leavevmode\hypertarget{ref-Fournier-Tondreau2024}{}%
Fournier-Tondreau, M., MacDonald, R. J., Radica, M., Lafrenière, D.,
Welbanks, L., Piaulet, C., Coulombe, L.-P., Allart, R., Morel, K.,
Artigau, É., Albert, L., Lim, O., Doyon, R., Benneke, B., Rowe, J. F.,
Darveau-Bernier, A., Cowan, N. B., Lewis, N. K., Cook, N. J., \ldots{}
Turner, J. D. (2024). {Near-infrared transmission spectroscopy of
HAT-P-18 b with NIRISS: Disentangling planetary and stellar features in
the era of JWST}. \emph{Monthly Notices of the Royal Astronomical
Society}, \emph{528}(2), 3354--3377.
\url{https://doi.org/10.1093/mnras/stad3813}

\leavevmode\hypertarget{ref-grant_david_2023_8211207}{}%
Grant, D., Valentine, D. E., \& Wakeford, H. R. (2023).
\emph{Exo-TiC/ExoTiC-MIRI: ExoTiC-MIRI v1.0.0} (Version v1.0.0)
{[}Computer software{]}. Zenodo.
\url{https://doi.org/10.5281/zenodo.8211207}

\leavevmode\hypertarget{ref-david_grant_2022_7437681}{}%
Grant, D., \& Wakeford, H. R. (2022). \emph{Exo-TiC/ExoTiC-LD: ExoTiC-LD
v3.0.0} (Version v3.0.0). Zenodo; Zenodo.
\url{https://doi.org/10.5281/zenodo.7437681}

\leavevmode\hypertarget{ref-harris2020array}{}%
Harris, C. R., Millman, K. J., Walt, S. J. van der, Gommers, R.,
Virtanen, P., Cournapeau, D., Wieser, E., Taylor, J., Berg, S., Smith,
N. J., Kern, R., Picus, M., Hoyer, S., Kerkwijk, M. H. van, Brett, M.,
Haldane, A., Río, J. F. del, Wiebe, M., Peterson, P., \ldots{} Oliphant,
T. E. (2020). Array programming with {NumPy}. \emph{Nature},
\emph{585}(7825), 357--362.
\url{https://doi.org/10.1038/s41586-020-2649-2}

\leavevmode\hypertarget{ref-Hunter:2007}{}%
Hunter, J. D. (2007). Matplotlib: A 2D graphics environment.
\emph{Computing in Science \& Engineering}, \emph{9}(3), 90--95.
\url{https://doi.org/10.1109/MCSE.2007.55}

\leavevmode\hypertarget{ref-Lim2023}{}%
Lim, O., Benneke, B., Doyon, R., MacDonald, R. J., Piaulet, C., Artigau,
É., Coulombe, L.-P., Radica, M., L'Heureux, A., Albert, L., Rackham, B.
V., de Wit, J., Salhi, S., Roy, P.-A., Flagg, L., Fournier-Tondreau, M.,
Taylor, J., Cook, N. J., Lafrenière, D., \ldots{} Darveau-Bernier, A.
(2023). {Atmospheric Reconnaissance of TRAPPIST-1 b with JWST/NIRISS:
Evidence for Strong Stellar Contamination in the Transmission Spectra}.
\emph{The Astrophysical Journal Letters}, \emph{955}(1), L22.
\url{https://doi.org/10.3847/2041-8213/acf7c4}

\leavevmode\hypertarget{ref-michael_radica_2024_12628066}{}%
Radica, M. (2024). \emph{Radicamc/exoUPRF: exoUPRF v1.0.0} (Version
v1.0.0) {[}Computer software{]}. Zenodo.
\url{https://doi.org/10.5281/zenodo.12628066}

\leavevmode\hypertarget{ref-Radica2022}{}%
Radica, M., Albert, L., Taylor, J., Lafrenière, D., Coulombe, L.-P.,
Darveau-Bernier, A., Doyon, R., Cook, N., Cowan, N., Espinoza, N.,
Johnstone, D., Kaltenegger, L., Piaulet, C., Roy, A., \& Talens, G. J.
(2022). {APPLESOSS: A Producer of ProfiLEs for SOSS. Application to the
NIRISS SOSS Mode}. \emph{Publications of the Astronomical Society of the
Pacific}, \emph{134}(1040), 104502.
\url{https://doi.org/10.1088/1538-3873/ac9430}

\leavevmode\hypertarget{ref-Radica2024}{}%
Radica, M., Coulombe, L.-P., Taylor, J., Albert, L., Allart, R.,
Benneke, B., Cowan, N. B., Dang, L., Lafrenière, D., Thorngren, D.,
Artigau, É., Doyon, R., Flagg, L., Johnstone, D., Pelletier, S., \& Roy,
P.-A. (2024). {Muted Features in the JWST NIRISS Transmission Spectrum
of Hot Neptune LTT 9779b}. \emph{The Astrophysical Journal Letters},
\emph{962}(1), L20. \url{https://doi.org/10.3847/2041-8213/ad20e4}

\leavevmode\hypertarget{ref-Radica2023}{}%
Radica, M., Welbanks, L., Espinoza, N., Taylor, J., Coulombe, L.-P.,
Feinstein, A. D., Goyal, J., Scarsdale, N., Albert, L., Baghel, P.,
Bean, J. L., Blecic, J., Lafrenière, D., MacDonald, R. J., Zamyatina,
M., Allart1, R., Artigau, É., Batalha, N. E., Cook, N. J., \ldots{}
Volk, K. (2023). {Awesome SOSS: transmission spectroscopy of WASP-96b
with NIRISS/SOSS}. \emph{Monthly Notices of the Royal Astronomical
Society}, \emph{524}(1), 835--856.
\url{https://doi.org/10.1093/mnras/stad1762}

\leavevmode\hypertarget{ref-tshirt2022}{}%
Schlawin, E., \& Glidic, K. (2022). Tshirt. In \emph{GitHub repository}.
GitHub. \url{https://github.com/eas342/tshirt}

\leavevmode\hypertarget{ref-2020SciPy-NMeth}{}%
Virtanen, P., Gommers, R., Oliphant, T. E., Haberland, M., Reddy, T.,
Cournapeau, D., Burovski, E., Peterson, P., Weckesser, W., Bright, J.,
van der Walt, S. J., Brett, M., Wilson, J., Millman, K. J., Mayorov, N.,
Nelson, A. R. J., Jones, E., Kern, R., Larson, E., \ldots{} SciPy 1.0
Contributors. (2020). {{SciPy} 1.0: Fundamental Algorithms for
Scientific Computing in Python}. \emph{Nature Methods}, \emph{17},
261--272. \url{https://doi.org/10.1038/s41592-019-0686-2}

\leavevmode\hypertarget{ref-pacman2022}{}%
Zieba, S., \& Kreidberg, L. (2022). PACMAN. In \emph{GitHub repository}.
GitHub. \url{https://github.com/sebastian-zieba/PACMAN}

\end{CSLReferences}

\end{document}